\documentclass[twocolumn,prl,superscriptaddress]{revtex4-1}
\usepackage{graphicx,subfigure,amsmath,amssymb,hyperref,mathrsfs}
\usepackage{setspace}

\usepackage{color}

\newcommand{\eps}[0]{\epsilon}

\newcommand{\beq}[0]{\begin{equation}}
\newcommand{\eeq}[0]{\end{equation}}
\newcommand{\beqa}[0]{\begin{align}}
\newcommand{\eeqa}[0]{\end{align}}

\def\be{\begin{equation}}
\def\ee{\end{equation}}
\def\e#1{\label{#1}\end{equation}}
\def\bea{\begin{eqnarray}}
\def\eea{\end{eqnarray}}
\def\ea#1{\label{#1}\end{eqnarray}}

\def\bem#1{\begin{mathletters}\label{#1}}
\def\eml{\end{mathletters}}

\def\4#1{{\boldsymbol{#1}}}
\def\8#1{{\widetilde{#1}}}
\def\bse{\begin{subequations}}
\def\ese{\end{subequations}}

\begin{document}

\title{Quantum bath refrigeration towards absolute zero: unattainability principle challenged.}

\author{M. Kol\'a\v r}
\thanks{Equal contribution}
\affiliation{Department of Optics, Palack\'{y}
University, 771 46 Olomouc, Czech Republic}
\author{D. Gelbwaser-Klimovsky}
\thanks{Equal contribution}
\affiliation{Weizmann Institute of Science, 76100
Rehovot, Israel}
\author{R. Alicki}
\affiliation{Weizmann Institute of Science, 76100
Rehovot, Israel}
\affiliation{Institute of Theoretical physics and Astrophysics,
University of Gda\'nsk}
\author{G. Kurizki}
\affiliation{Weizmann Institute of Science, 76100
Rehovot, Israel}

\begin{abstract}
A minimal model of a quantum refrigerator (QR), i.e. a periodically phase-flipped two-level system permanently coupled to a finite-capacity bath (cold bath) and an infinite heat dump (hot bath), is introduced and used to investigate the cooling of the cold bath towards the absolute zero ($T=0$). Remarkably, the temperature scaling of the cold-bath cooling rate reveals that it does not vanish as $T\rightarrow0$ for certain realistic quantized baths, e.g. phonons in strongly disordered media (fractons) or quantized spin-waves in ferromagnets (magnons). This result challenges Nernst's third-law formulation known as the unattainability principle.
\end{abstract}
\maketitle

{\em Introduction.} One of the generally unsettled fundamental problems  of  thermodynamics is the nature of  the ultimate limitations on cooling to absolute zero, $T=0$. Attaining $T=0$ in a finite number of steps, or, more generally,\textit{ in finite time}, is prohibited by Nernst's \textit{unattainability principle} which is the \textit{dynamical} formulation of the third law of thermodynamics \cite{nernst,landsberg,belgiorno}.However, the universality of this principle has been postulated rather than proven. It is also debatable whether this formulation is always equivalent to Nernst's \textit{heat theorem}, whereby the \textit{entropy vanishes} at $T=0$.Do both formulations of  the third law hold for all \textit{quantum} scenarios?
The investigation of this open fundamental problem, pertaining to  quantum refrigerator (QR) schemes 
\cite{gordon-jap,segal-pre,linden,kosloffprl12,SchulzePRA10}, 
raises several  principal questions: (i) How does the \textit{cooling rate} scale with the bath temperature and does it necessarily vanish as $T\rightarrow0$? (ii) Does a QR differ from its classical counterpart regarding compliance with the third law and to what extent is such compliance model-dependent? Answers to these questions are important not only to the understanding of the foundations of quantum thermodynamics \cite{alicki}, but also to the design of novel QR schemes compatible with the needs of quantum nanotechnologies %\cite{wrachtrup,schmiedmayer} 
that require compact (nanosize) coolers capable of ultrafast cooling \cite{pop}.  

Here, we propose and explore the simplest QR design thus far   that allows us to address the fundamental issues raised above.
The working medium is a \textit{single two-level system (qubit), permanently} (rather than intermittently, as is done in traditional cycles \cite{gordon-jap}) coupled to a finite-capacity  bath to be cooled and to another,  much larger and hotter, heat dump. The pumping operation consists of fast modulation of the qubit energy by means of periodic $\pi$-flips of the qubit phase.
We find that this QR  can cool down a finite-capacity (yet macroscopic and spectrally-continuous) bath only if the modulation-period  is within the bath-memory (non-Markovian) time. Hence, the cold-bath spectrum is crucial in determining the cooling condition and rate. Our most striking finding is that for certain experimentally realizable baths, such as quantized spin-waves in ferromagnets (magnons) \cite{kittel} or acoustic phonons in strongly disordered media (fractons) \cite{Alexander}, the cooling rate  remains finite as $T\rightarrow0$, in apparent violation of  the dynamical  formulation of the third law. 

{\em Model and analysis}. A control qubit is weakly coupled to two  baths via the system-bath interaction hamiltionan: $H_{SB}=\sigma_x(B_H+B_C)$, where $\sigma_x$ is the spinor x-component,$B_C$ is the operator of a finite cold bath ( C) which we wish to refrigerate, and $B_H$ that of a much larger hot bath ( H) that remains nearly unchanged.  The qubit energy is periodically modulated by an external field $\nu(t)$ via the Hamiltonian $H_{ext}=\frac{1}{2}\sigma_z \nu(t)$.  An illustration (Fig. 1-inset) is that of a charged quantum particle in a double-well potential  that is periodically phase-flipped by off-resonant pulses and is coupled to a spatially-confined (macroscopic) C-bath to be cooled, as well as to a nearly-infinite H environment into which the heat is dumped. This scheme bears analogy to radiative (sideband) cooling in solids and molecules\cite{R1}, if one visualizes the red- and blue- shifted qubit frequencies as Stokes and anti-Stokes lines respectively.

 Our analysis reveals the crucial role of the quantized characteristics of system-bath coupling in determining the attainability of $T\rightarrow0$. By contrast, the results are insensitive to the QR scheme chosen (see Discussion).

 The general condition for steady-state refrigeration, under periodic, off resonant, modulation, is \textit{positive}  heat current from C to H via the qubit. The sign and magnitude of the current  is determined  by the  steady-state solution of a non-Markovian master equation (ME) for the qubit density matrix \cite{kurizki-prl}. 
 The ME, which is accurate to second order in the system-bath coupling, allows for  time-dependent modulation of the system that is  much faster than the bath-memory time $t_c$. It is  valid at any T, as shown both theoretically \cite{kurizki-prl} and experimentally \cite{Almog,alvarez}. Deviations of the evolution (Suppl. A) from that described by the non-Markovian ME include  system-bath entanglement (correlations) effects that can be compensated by readjusting the qubit excitation, as well as  bath dynamics effects (violation of the Born approximation whereby the bath is constantly in a thermal state)\cite{noam-nature}. Yet such deviations    are of \textit{fourth-order} in the system-bath coupling and thus negligible for weak coupling. The cooling of a finite-capacity bath is the result of infinitesimal temperature changes over many modulation cycles, consistently with the Born approximation underlying the ME. Since the Born approximation is the more accurate the larger the bath\cite{17}, we assume that the finite-capacity bath is macroscopic and has a continuous spectrum, which does not exhibit mode discreteness or recurrences that may otherwise invalidate this approximation and bath thermalization altogether \cite{17}.
 
 \textit{Only the  diagonal elements} of the qubit's density matrix $\rho_S$ (energy-state populations) play a part here, although the ME  also allows for  coherences (off-diagonal elements)[14], but these are   absent at $t=0$ (starting at equilibrium) and remain so under the modulation. The quantumness of the ME, even when it is diagonal in the energy basis, is embodied by the qubit interlevel transition rate and  their non-Markovian time-dependence  (Suppl. B).
Periodic phase shifts of the qubit at intervals $\tau$ dynamically control its coupling to the baths and the resulting transition rates. When $\tau$ is  comparable to the bath memory-time $t_c$, these phase shifts  modify the detailed balance of the transition rates and thereby either heat or cool the qubit depending on $\tau$ \cite{alvarez,noam-nature}.
In what follows, we analyze the steady state and  the slow changes of the \textit{bath} temperature as a result of these periodic perturbations.

Under weak-coupling conditions, the qubit evolution caused by  the baths is much slower than $\tau \sim t_c$. Hence, in steady state,  we can use \textit{time-averaged   level populations and transition rates} between the periodically-perturbed  qubit levels (Suppl. B).
These time-averaged (steady-state) equations  can be recast, upon introducing the polarization of the qubit $S\equiv (\rho_{ee}-\rho_{gg})/2$,   into 
\begin{eqnarray}
\dot{\overline{S}}=-\left[\overline{R_g}+\overline{R_e}\right]\overline{S}+\frac{\overline{R_g}-\overline{R_e}}{2},
\label{master-eq}
\end{eqnarray}
Here the $\left|e\right>\rightarrow \left|g\right>$  averaged transition rate from the excited (e) to the ground (g) state, is $\overline{R_e}$ and its $\left|g\right>\rightarrow \left|e\right>$ counterpart is $\overline{R_g}$. 
The averaged transition rates for $t\gg\tau$  are found, 
upon expanding the   qubit energy  under periodic frequency modulation $\nu(t)$ into the harmonic (Floquet) series (Suppl. B, \cite{tutorial})

\begin{multline}
\overline{R}_{e(g)}\equiv 
2\pi\sum_m P_m G_T[\pm(\omega_0+m\Delta)];\\
P_m = |\varepsilon_m|^2\  , \ \varepsilon_m = \frac{1}{\tau}\int_0^{\tau} e^{i\int_0^t(\nu(t')-\omega_0)dt' }e^{im\Delta t} dt,
\label{decay-rates-averaged-def}
\end{multline}

\noindent Here m are all (positive and negative) integers,  $P_m$ are the probabilities of shifting  $G_T(\omega)$  by $m\Delta$, $\Delta=\frac{2\pi}{\tau}$, from the qubit average frequency $\omega_0$,  $G_T(\omega)$ being  the temperature-dependent bath-coupling spectrum, i.e. the Fourier transform of the bath autocorrelation function:
$
G_T(\omega)= \int_{-\infty}^{+\infty} e^{i\omega t}\langle B(t)B(0)\rangle dt = e^{\omega/k_B T} G_T(-\omega) .
$
For a bosonic bath:

\begin{eqnarray}
G_T(\omega)=&G_0(\omega)(n(\omega)+1);  \hspace{0.5cm}  G_0(\omega)=|g(\omega)|^2\rho(\omega); \notag \\
 n(\omega)=\frac{1}{e^{\frac{\omega}{T}}-1},\notag\\
  \label{eq:G}
 \end{eqnarray}
 $g(\omega)$ being the system-bath coupling, $\rho(\omega)$ the bath -mode density and $n(\omega)$ the $\omega$-mode thermal occupancy. These expressions are also obtainable by Floquet (harmonic) expansion of the periodically-driven Markovian (Lindblad) Liouvillian \cite{tutorial,alilid}.

In the presence of hot ($H$) and cold ($C$) baths, under the assumption of a qubit weakly coupled to both baths,  the transition rates are split into {\em additive harmonic contributions}: 
$\overline{R_{e(g)}}\equiv \sum_m \left( \overline{R_{e(g)}^{C(m)}}+\overline{R_{e(g)}^{H(m)}}\right)$.
Hence, Eq.(1) is also split into (Suppl. B) $\dot{\overline{S}}=\sum_m(\dot{\overline{S^C_m}}+\dot{\overline{S^H_m}})$, where $\dot{\overline{S^{C(H)}_m}}$ is the m-harmonic {\em polarization} flow caused by the cold (hot) bath only.
The averaged heat flow,$\dot{\bar{Q}}$, through the qubit  is correspondingly divided into the $C$ and $H$ bath-contributions. The steady-state Eq. (1) then gives rise to 
\begin{equation}
J_{C(H)}=\dot{\overline{Q}}_{C(H)}=\sum_m(\omega_0+m\Delta)\dot{\overline{S_m^{C(H)}}}
\label{eq:currentheat}
\end{equation}

which is the sum of rates of heat-exchange with the respective  baths via $\omega_0+m\Delta$ quanta.  Positive $J_C$ implies refrigeration, i.e., \textit{heat flow from the cold bath to the hot bath} via the modulated qubit.

It is advantageous to use periodic, alternating, $\pi$-phase shifts (phase flips) as they give   rise, to leading order, to two symmetrically opposite frequency shifts  of $G_T$ at  $\omega_0\pm\Delta$.  Then, by the Floquet expansion we obtain the probability distribution 
  wherein $P_0=0$ and $P_{\pm 1}\approx (2/\pi)^2$ are the leading terms \cite{kurizki-prl,Kurizki:2001}.

  Let us choose sufficiently large $\Delta$, of the order of the  spectral width $\Gamma=1/t_c$, which is the inverse memory time of the cold bath, such that at $\omega \simeq \omega_0+\Delta$ the qubit is  coupled only to the hot bath, while at $\omega\simeq\omega_0-\Delta$ it is  coupled to both the cold  and the hot baths. 
More precisely, we require that 
\begin{eqnarray}
G_T(\omega_0+\Delta)\approx  G_T^H(\omega_0+\Delta)\gg G_T^C(\omega_0+\Delta); \notag \\
 G_T^H(\omega_0+\Delta)\gg G_T^H(\omega_0-\Delta), G_T^C(\omega_0-\Delta)
\label{eq:gt}
\end{eqnarray}

where $G_T^{C(H)}(\omega)$ is the respective temperature-dependent bath-coupling spectrum.
This requirement can be satisfied if the cold bath ($C$) is spectrally  localized  with upper cutoff $\omega_{\rm cut}<\omega_0+\Delta$. By contrast, for the hot bath ($H$), the required rise  of $G_T^H$ with $\omega$ is obtained for most common bath spectra, provided the cutoff of $G_T^H(\omega)$ is much higher than $\omega_{cut}$ of $G_T^C(\omega)$: e.g.,  for phonons in bulk media or photons in open space, $G_T^H(\omega)\propto \omega^3$ satisfies Eq. \eqref{eq:gt}.

%For $T_H>T_C$, $n_T^{C}(\omega_0)<n_T^{H}(\omega_0) $.  
\begin{figure}
	\centering
		\includegraphics[width=0.8\linewidth]{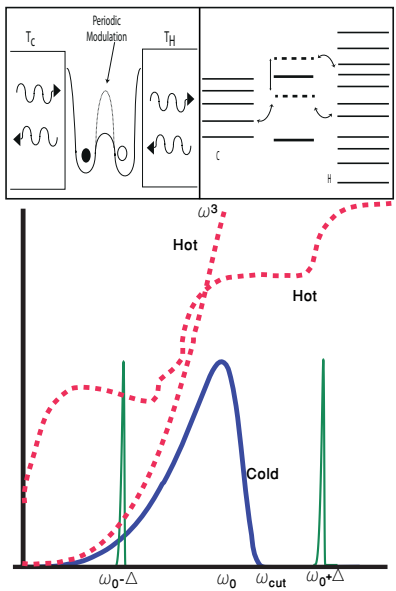}
	\caption{Main panel: Schematic depiction of the required bath spectra and the qubit frequency shifts due to periodic phase flips. Inset: Schematic realization of the modulated qubit and its coupling to the baths.}
	\label{fig:bathspectrum}
\end{figure}

Under the conditions of   Eq. \eqref{eq:gt} we find  that  the steady-state heat current from $C$ to $H$ is (Suppl. B)
\begin{eqnarray}
J_C&=&(\omega_0-\Delta)\dot{\overline{S^C_S}}=\\
\nonumber
&\simeq&(\omega_0-\Delta)\frac{G_0^C(\omega_0-\Delta) [n^C(\omega_0-\Delta)-n^H(\omega_0+\Delta)]}{[2n^H(\omega_0+\Delta)+1]}.
\label{eq:coldcurrent}
\end{eqnarray}
The balance of $J_C$ and $J_H$ (cold and hot ) currents \eqref{eq:currentheat}  \textit{obeys  the second law} \cite{landsberg,lindblad}: It can be verified that the entropy production rate $\frac{d\mathscr{S}}{dt}$ satisfies:
$
\frac{d\mathscr{S}}{dt}-
\left( \frac{J_C}{T_C}+\frac{J_H}{T_H} \right)\geq0
$
for any initial state.

From Eq. (6), the heat pump (QR) condition $ J_C>0$ amounts to 
\begin{equation}
n^C(\omega_0-\Delta)>n^H(\omega_0+\Delta)\Leftrightarrow
\frac{\omega_0+\Delta}{T_H}>\frac{\omega_0-\Delta}{T_C}.
\label{eq:condflat}
\end{equation}

\noindent An analogous relation holds if $n^{C(H)}(\omega)$ are Boltzmann rather than Bose factors (occupancies).

Equation \eqref{eq:condflat} reveals the crux of the heat pumping (QR) effect: although by definition $n^C(\omega_0)<n^H(\omega_0)$, heat can flow from the cold  to the hot   bath if the $C$-bath thermal occupancy at $\omega_0-\Delta$ is higher than that of the $H$-bath at $\omega \simeq \omega_0+\Delta$.  
If $\Delta$ is too small for Eq.~\eqref{eq:condflat} to hold, we recover the natural heat flow direction $H\rightarrow S \rightarrow C$ at steady-state.
In addition, Eq.\eqref{eq:gt} implies that the heat pump requires the qubit to be {\em simultaneously} coupled to the $C$ and $H$ baths at $\omega_0-\Delta$ and $\omega_0+\Delta$, respectively. 
\par
{\em Cooling rate scaling with temperature}. In what follows, we investigate the QR action (heat pumping from $C$ to $H$) under the  assumptions that the hot bath is practically infinite, hence $T_H={\rm const}$, whereas  the macroscopic cold bath has \textit{finite} {\em heat capacity}, $c_V<\infty$, resulting in  \textit{slow evolution} of $T_C(t)$  under the QR action.
To estimate this evolution  we use the standard thermodynamic definition \cite{reichl}
\begin{equation}
c_V\,\frac{{\rm d}T_C(t)}{{\rm d}t}=J_C=\dot{\bar{Q}}_C
\label{temperature-equation}
\end{equation}
which presumes that $T_C$ is well-defined at all t (since the bath has a continuous spectrum and is large enough to thermalize at finite times).

In order to infer the temperature-dependence of the cooling rate $\frac{dT_C}{dt}$ we shall examine the scaling of $c_V$ and $J_C$ with $T_C:$
a)The constant-volume heat capacity of the cold bath, $c_V$,  depends on the dimensionality of the bosonic bath. If $\rho(\omega)\simeq\omega^{d-1}$ is the $d$-dimensional density of modes and $T_C\ll \omega_{cut}$ ($k_B=\hbar=1$), then
\begin{multline}
\lim_{T_C\rightarrow 0}c_V=\\
\frac{d}{d T}
 \frac{\langle H_B \rangle}{V}|_{T_C}
 \simeq \frac{d}{d T}
 \int d\omega \omega \rho(\omega) 
 (n_C(\omega)+1)|_{T_C} \sim T_C^d
\label{heat-capacity}
\end{multline}

b) The scaling of the cold-bath heat current, $J_C$, in Eq. \eqref{eq:coldcurrent}  can be deduced if we maximize the heat flow \cite{jens}  with respect to $\Delta$ (our control parameter). This gives  the dependence of $\omega_0-\Delta \simeq T_C$ \cite{kosloffprl12,arxivekos}.
Hence, to maintain the maximum heat flow, we have to slowly increase $\Delta$ with time, so as to approach $T_C\rightarrow 0$.
The closer to $T_C\rightarrow 0$, the lower is $\omega_0-\Delta$, hence the steady-state dynamics \eqref{eq:coldcurrent} and its slow change \eqref{temperature-equation} become increasingly more accurate. Correspondingly, we parametrize $J_C$  in Eq. (6) using Eq. (3) and assuming the low-frequency range of the cold bath $0\leq\omega=\omega_0-\Delta\ll\omega_{cut}$:
$
\lim_{\omega\rightarrow 0}|g(\omega)|^2\propto  \omega^\gamma,\;\rho(\omega)\approx\omega^{d-1}.
$
%where for $0<\alpha<1$ ($\alpha>1$) the bath is termed sub-Ohmic (super-Ohmic), respectively, $A$ is a constant, 
Here, $|g(\omega)|^2$ is the $\gamma$- dependent system-coupling to the bosonic bath (discussed below). The heat current, maximized for $\omega_0-\Delta\approx T_C$, then obeys  the scaling
\begin{equation}
J_C(T_C)\propto -T_C^{\gamma+d}.
\label{JCapproximation-general}
\end{equation}

c) Upon substituting Eqs. \eqref{heat-capacity}-\eqref{JCapproximation-general} in Eq. \eqref{temperature-equation} we observe that the $T_C^d$ scaling of $c_V$ is canceled by a similar scaling of the density of modes in Eq. \eqref{JCapproximation-general}. The resulting scaling  yields
%of Eq.~\eqref{temperature-equation}  
\begin{equation}
{\rm d}T_C/{\rm d}t=-A T_C^\gamma. 
\label{tc-scaling-general}
\end{equation}

\noindent Here the constant $A\propto 1/V$: the larger the bath the slower its cooling. 

\begin{figure}
	\centering
		\includegraphics{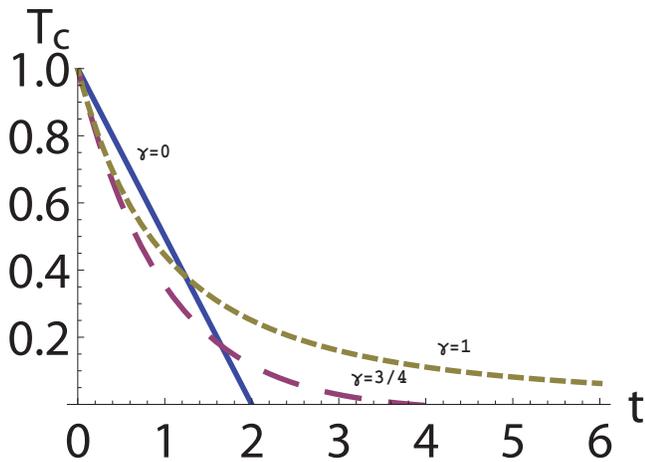}
	\caption{$T_C$ change with time (cooling) for three different system-bath coupling-strength dispersion laws: $\gamma=1$ (acoustic phonons), $\gamma=3/4$ (fractons), $\gamma=0$ (magnons).}
	\label{with-modulation20modes}
\end{figure}

Remarkably,  the $\frac{dT_C}{dt}$ scaling only depends on the $\gamma$-th scaling power of the system-bath coupling strength $|g(\omega)|^2$.
For different $\gamma$ the time-dependence of $T_C$, starting from the same $T_C(0)$, is plotted in Fig. \ref{with-modulation20modes}. 
For $\gamma =1$ we have exponentially slow convergence to $T_C\rightarrow 0$, conforming to the third law. Yet, strikingly, for $0\leq\gamma<1$, $T_C(t)\rightarrow 0$ at {\em finite} time, thus violating the accepted dynamical  formulation of the third law \cite{nernst,landsberg,belgiorno}, if the frequency-dependent coupling $|g(\omega)|^2$ is {\em sub-linear} in $\omega$. 

In what follows, we examine the possibility of such scaling for different bosonic baths. To this end, consider a qubit immersed in a periodic medium, whose local displacement  is a  linear combination of normal-mode  creation and  annihilation  operators (bath excitations/de-excitations)
$\hat{B}(\vec{\mathbf{x}}) = \frac{1}{\sqrt{V}}\sum_{\vec{\mathbf{k}}}\frac{1}{\sqrt{\omega(\vec{\mathbf{k}})}}\bigl(\phi_{\vec{\textbf{k}}}(\vec{\textbf{x}}) a^{\dagger}(\mathbf{k}) + h.c.\bigr)
$.
The normal-mode functions are labeled by the wave vectors $\mathbf{k}$ that  belong to a reciprocal lattice bounded by the Debye cutoff $(\omega(\mathbf{k})\leq \omega_{cut}=\omega_D)$. The couplings  of a charged or dipolar system to bath excitations/deexcitations are to leading order determined by the gradient of the displacement operator
\begin{equation}
\nabla\hat{B}(\vec{\mathbf{x}}) =
 \frac{-i}{\sqrt{V}}
 \sum_{\vec{\textbf{k}}}
 \frac{1}{\sqrt{\omega(\vec{\mathbf{k}})}}
 \bigl( \nabla \phi_{\vec{\textbf{k}}}(\vec{\textbf{x}}) a^{\dagger}(\vec{\mathbf{k}}) - h.c.\bigr)
\label{grad}
\end{equation}

\noindent When $ \phi_{\vec{\textbf{k}}}(\vec{\textbf{x}})=e^{-i\vec{\textbf{k}}\cdot\vec{\textbf{x}}}$ the corresponding coupling constant scales as $|g(\omega(\vec{\textbf{k}}))| \sim \frac{\vec{\textbf{k}}}{\sqrt{\omega(\vec{\textbf{k}})}}$.  We can discern  three generic types of scaling of the coupling constant: 

i)For \textit{acoustic phonons}  $\omega(\vec{\mathbf{k}}) \simeq v|\vec{\mathbf{k}}|$, where $v$ is a sound velocity and the coupling strength  satisfies $|g(\omega)|^2\sim \omega $, i.e. $\gamma = 1$. Therefore,  acoustic phonons used as a cold bath do not violate the dynamical third law formulation: the optimal cooling to zero temperature is \textit{exponential in time}.
\par
ii) Amorphous (glass) materials  may exhibit  effects of  fractal  disorder. These effects imply  different scaling of the displacement of the mode function $\phi_{\vec{\textbf{k}}}(\vec{\textbf{x}}),$
$|\nabla\phi_{\vec{\textbf{k}}}(\vec{\textbf{x}})|\sim \omega^\gamma |\phi_{\vec{\textbf{k}}}(\vec{\textbf{x}})|
$:  normal phonons   are replaced by  \emph{fractons}  for which $\gamma$ takes fractional values. In particular, for some materials $\gamma <1$ \cite{Alexander}. Hence, for a cold bath composed of such fractons  the  violation of the third law is expected.
\par
iii) Another system which leads to a  violation of the third law is the magnon (spin-wave) bath in a ferromagnetic spin lattice with nearest neighbor interactions,  below the critical temperature. 
The Holstein-Primakoff transformation of the jth spin  Pauli matrix \cite{kittel}, 
$
S^+_{j} = S_{jx} + i S_{jy} = (2S)^{1/2}\bigl( 1- a^{\dagger}_j a_j/2S\bigr)^{1/2} a_j \notag
$ to boson annihilation and creation operators $a_j , a^{\dagger}_j$, allows to represent the system as a set of interacting  harmonic oscillators. Introducing the collective spin-wave  (magnon) variables $a(\vec{\mathbf{k}}) , a^{\dagger}(\vec{\mathbf{k}})$ satisfying
$
a_j  = \frac{1}{\sqrt{N}}\sum_{\vec{\mathbf{k}}} e^{-i\vec{\mathbf{k}}\cdot\vec{\mathbf{x}}_j}a(\vec{\mathbf{k}}),
$%
 we can rewrite its Hamiltonian  in the form
$H_0 = \sum_{\vec{\mathbf{k}}} \omega(\vec{\mathbf{k}}) a^{\dagger}(\vec{\mathbf{k}})a(\vec{\mathbf{k}})+Higher\, order\, terms$.
 At low temperatures  the nonlinearity in the Holstein-Primakoff transformation can be neglected
and  the system becomes equivalent to a bosonic system governed by  the Hamiltonian $H_0$, whereby
the dispersion law is quadratic in the low-frequency region, $\omega(\mathbf{k})\sim (|\mathbf{k}|^2 + \mathrm{constant})$.  The local spin variable $a_j$ can then  be \textit{directly coupled} to the qubit  by a dipole-dipole (spin-spin) interaction. Hence, the main difference between the dipolar coupling to acoustic phonons and magnons is the absence of the dispersive-coupling coefficient $\frac{\vec{\mathbf{k}}}{\sqrt{\omega(\vec{\mathbf{k}})}}$ for the latter. Therefore, the coupling strength  to magnons satisfies $|g(\omega)|^2\sim 1$ ($\gamma = 0$), which implies the  violation of the third law for magnons.

{\em Discussion.}
We have analyzed the  cooling process of a bosonic bath towards the absolute zero using a new minimal model  of a  quantum refrigerator: a single two-level system (qubit) {\it permanently} coupled to a spectrally-restricted cold bath with finite heat capacity and a  hot bath with infinite heat capacity  has been shown to act  as a heat pump, under appropriate modulation. The heat flow is proportional to the population-difference of a pair of oppositely-shifted  bath modes that are selected by the qubit modulation (phase-flip) rate, analogously to sideband cooling \cite{R1}. The attainable cooling rate challenges the third law of thermodynamics, in the sense that  arbitrarily low temperature of the cold bath may be reached in {\em finite} time by the\textit{ heat pump} for certain \textit{quantized} cold-bath spectra: e.g. magnon and fracton baths. 

 In solid-state ferromagnets or glasses, interactions  of control qubits with other baths unaccounted by the model, as well as tiny deviations from the predicted weak-coupling, steady-state dynamics  (discussed in Suppl. A,B ) may restore the third law. Nevertheless, \textit{surprisingly fast cooling} ($\gamma<1$) may still be observed down to some (material-dependent) temperature. It would be preferable to demonstrate this effect for quantum dots coupled to  controllable baths composed of nuclear spins in solids \cite{NeumannSCI08} or for atomic dipoles in optical lattices \cite{diehl}: in both cases the systems  are highly shielded from other baths, while the lattices can be engineered to conform to the nearest-neighbor ferromagnetic model that engenders magnons.

This  study of the compliance with the third law for {\em quantized non-Markovian baths} indicates that the temperature scaling of the cooling rate is not specific to the chosen QR model; it is similar to the scaling obtained for the very different noise -driven QR\cite{arxivekos}. Namely, the scaling is not sensitive to the form of driving, nor to the method of treating the steady-state dynamics. Hence, the dependence of the scaling on the  \textit{system-bath coupling dispersion} is general. It
 provides new insights into the bounds  of bath cooling  in quantum thermodynamics.  It shows that Nernst's  principle of unattainability of the absolute zero in finite time \cite{nernst,landsberg,belgiorno} may fail and is not always equivalent to Nernst's heat theorem (see Introduction): the latter holds true since a bosonic bath has a unique ground state whose entropy must vanish.

\textit{Acknowledgements} The support of EC(FET Open), DIP, the Humboldt-Meitner Award(G.K.), the Weston Visiting Professorship (R.A.), CONACYT (D.G.) are acknowledged. M.K. the Czech Science Foundation, GAP205/10/1657 (M.K.). Useful discussions are acknowledged with Nir Gov and Ronnie Kosloff

\section{Supplementary A: Universality and Accuracy of the Non-Markovian Master equation}
\setcounter{equation}{0}
\renewcommand{\theequation}{A\arabic{equation}}

The dynamics derived in the main text can be formulated for the following  Hamiltonian for a phase-modulated qubit (TLS) weakly coupled to a general bath:
\bea
\label{H_tot}
H_{tot} &=&H_S+H_B + \epsilon H_{SB},\\
H_S &=& \frac{\nu(t)}{2}\sigma_z\\
\label{H_SB}
H_{SB}&=&\sigma_xB,
\eea
where $H_B$ is any time-independent bath Hamiltonian and $B$ is any time-independent bath operator and

\begin{equation}
\epsilon = \max(\eta_k/\omega_0),
\end{equation}

 is a small dimensionless parameter normalizing the rate $\eta_k$ of the maximally coupled bath mode divided by the TLS resonant frequency, $\omega_0$.
 
As shown in Ref.\cite{kurizki-prl}, one can derive, to second order in the system-bath coupling, the following non-Markovian master-equation for the reduced system density matrix, $\rho_S(t)$:
\begin{multline}
\label{gen-ME}
\dot{\rho_S}(t) = \\
-i\left[H_S,\rho_S(t)\right]+\\
\int_0^t d\tau \left\{
\Phi_T(t-\tau) \left[\tilde{S}(t,\tau)\rho_S(t),\sigma_x\right] +H.c.
\right\}
\end{multline}
where
\be
\Phi_T(t) = \epsilon^2\langle B e^{-iH_Bt}Be^{iH_Bt} \rangle_B,
\ee
is the correlation function of the bath and
\be
\tilde{S}(t,\tau) = e^{-iH_S(t-\tau)}\sigma_xe^{iH_S(t-\tau)}.
\ee

If we start at equilibrium, where $\rho_S$ is diagonal in the energy basis of the TLS ($|e\rangle$, $|g\rangle$) then it remains so under the action of the diagonal $H_S(T)$. This ME does not invoke the rotation-wave approximation \cite{kurizki-prl,noam-nature} and hence allows for arbitrarily fast modulations of the system, which cause its anomalous heating or cooling on time scales comparable to $\omega_0^{-1}$. The resulting rate equations are then given  by
\be
\label{eq-Bloch}
\dot\rho_{ee}(t) = -\dot\rho_{gg}(t) = R_g(t)\rho_{gg}-R_e(t)\rho_{ee},
\ee
where the non-Markovian time-dependent rates  $R_{g(e)}(t)$ are given for periodic  phase flips in Suppl. B.
It yields a Floquet expansion of $\rho_{ee}(t)$ and the corresponding polarization, as detailed in Suppl. B.

\subsection*{Accuracy of the Non-Markovian Master Equation}
The second-order  (Zwanzig-Nakajima type \cite{kurizki-prl}) non-Markov Master Equation (ME) involves two crucial
approximations:
The first is that the initial state is a product state of the qubit and bath density
matrices. The second is that the bath state $\rho_B$ does not change during the evolution
(Born approximation).
These approximations yield:
\begin{eqnarray}
\label{eq23}
\rho_{tot}(t)\Big{|}_{\rm ME}=\rho_S(t)\otimes\rho_B\Big|_{\rm ME}.
\end{eqnarray}
There are two types of deviations of the ME-described initial state from its exact (initial
state) evolution.

\begin{enumerate}
\item The first results from the system-bath correlation  effect that
does not enter the second-order ME calculation.  This deviation raises
the effective qubit temperature above that of the bath, by an extra
excitation,
\beq
 \delta\rho_{ee}= \frac{1}{2}(\langle \sigma_z(\epsilon)\rangle-\langle\sigma_z(\epsilon=0)\rangle)\sim O(\epsilon^2).
\eeq
It can \textit{be compensated }upon using the exact state populations \cite{noam-nature}:
\begin{eqnarray}
\label{delta-rho}
\rho_{ee}(t)=\rho_{ee}(t)\Big{|}_{\rm ME}+\delta\rho_{ee},\quad\rho_{gg}(t)=1-\rho_{ee}(t).
\end{eqnarray}

This correction vanishes over time intervals $\geq \omega_0^{-1}$ and does not affect the steady-state polarization (heat-flow) rates considered in Suppl. B.
\item
The other type of deviations is due to the bath: the ME assumes that $\rho_B$ is in a Gibbs (thermal) state, while the exact state has small deviations from it, responsible for  correlations among the bath modes. These are, in general, complicated functional forms of temperature and frequencies (see  Suppl. Material of \cite{noam-nature}, particularly  Gordon et al., (NJP \textbf{11}, 123025(2009)). The deviation from the Gibbs state used in~\eqref{eq23} is only  of fourth order in $\epsilon$, which can be neglected in the weak-coupling limit.  
\end{enumerate}
More generally, we can show that the deviations from master equation analysis are always to fourth order in $\epsilon$. To this end, the total
density matrix (of the system+bath) at any  time $t$ can be expanded in powers of $\epsilon$ as
\begin{eqnarray}
\label{eq28}
\rho_{tot}(t)=\rho^{(0)}_{tot}(t)+\eps\rho^{(1)}_{tot}(t)+\eps^2\rho^{(2)}_{tot}(t)+\cdots.
\end{eqnarray}
where $\rho^{(i)}_{tot}(t)$ need not necessarily be a \textit{valid} density matrix. Upon inserting~\eqref{eq28} in the equation of motion 
for $\rho_{tot}(t)$, and collecting terms of successive orders of $\epsilon$, we get
\begin{multline}
\label{eq29}
\dot\rho_{S} =\\
-i\eps Tr_B[H_{SB}(t),\rho_{tot}(0)]-\\
\eps^2\int_0^tdt'Tr_B[H_{SB}(t),[H_{SB}(t'),\rho_{tot}(t)]] + O(\epsilon^4),
\end{multline}

where the time-dependence of $H_{SB}(t)$ is determined by $H_S(t)$. It is important to note that odd powers of $\epsilon$ do not contribute, since the trace of the interaction Hamiltonian $H_{SB}$ vanishes.

One can see upon collecting the even powers of $\eps$ in Eqs. \eqref{eq28}, \eqref{eq29}, that the corrections to the master-equation analysis are always of fourth order in $\epsilon$.

\section*{Supplementary B: Floquet-expansion of polarization and transition rates}
\label{SB}
\setcounter{equation}{0}
\renewcommand{\theequation}{B\arabic{equation}}
%\section{Supplementary information}
%\subsection{ General expressions }
Under energy  modulation and in the presence of weak coupling to two  baths,  the non-Markovian ME leads to the following additive contributions to the  expressions for qubit polarization
\begin{multline}
\dot{S}=\dot{S}^C+\dot{S}^H=\\
 -(R_g^C+R_e^C)S+\frac{R_g^C-R_e^C}{2}-(R_g^H+R_e^H)S+\frac{R_g^H-R_e^H}{2}.
\end{multline}

From Ref.~\cite{kurizki-prl} we have the $|e\rangle \rightarrow |g\rangle$ and $|g\rangle \rightarrow |e\rangle$ transition rates, respectively, as
\begin{multline}
R_e(t)=
2\,{\rm Re}\!\int_0^t{\rm d}t^\prime\exp[i\omega_0(t-t^\prime)]\varepsilon(t)\varepsilon^\star(t^\prime)\Phi_T(t-t^\prime),\label{rates-def}\\
R_g(t)=2\,{\rm Re}\!\int_0^t{\rm d}t^\prime\exp[-i\omega_0(t-t^\prime)]\varepsilon^\star(t)\varepsilon(t^\prime)\Phi_T(t-t^\prime),
\end{multline}
Here ``Re'' stands for real part, $\varepsilon(t)$ for the modulated phase factor (a unimodular periodic complex function)  $\omega_0$ is the qubit's resonance frequency and the bath-correlation function $\Phi_T(t)\equiv \int {\rm d} \omega G_T(\omega)\exp(-i\omega t)$ is the  Fourier transform of to  the bath coupling spectrum $G_T(\omega)$.

The Floquet expansion of $\varepsilon(t)$ can be substituted into Eq. \eqref{rates-def}, to obtain ($k,l$ being integers)

\begin{multline}
\nonumber
R_e(t)=2\,{\rm Re}\!\int_0^t{\rm d}t^\prime\exp[i\omega_0(t-t^\prime)]\varepsilon(t)\varepsilon^\star(t^\prime)\Phi_T(t-t^\prime)\\
\nonumber
=2\int_{-\infty}^\infty{\rm d}\omega G_T(\omega)\\
\sum_{k,l}\varepsilon_k \varepsilon_l^\star
\left\lbrace \frac{\cos[(\omega_k-\omega_l)t]\sin[(\omega_0+\omega_l-\omega)t]} {\omega_0+\omega_l-\omega}\right.\\
\nonumber
\left. -\frac{2\sin[(\omega_k-\omega_l)t]\sin^2[(\omega_0+\omega_l-\omega)t/2]} {\omega_0+\omega_l-\omega}\right\rbrace,   \\
\nonumber
R_g(t)=2\,{\rm Re}\!\int_0^t{\rm d}t^\prime\exp[-i\omega_0(t-t^\prime)]\varepsilon^\star(t)\varepsilon(t^\prime)\Phi_T(t-t^\prime)\\
\nonumber
=2\int_{-\infty}^\infty{\rm d}\omega G_T(\omega)
\\\sum_{k,l}\varepsilon_k^\star \varepsilon_l\left\lbrace \frac{\cos[(\omega_k-\omega_l)t]\sin[(\omega_0+\omega_l+\omega)t]} {\omega_0+\omega_l+\omega}  \right.\\
\left. -\frac{2\sin[(\omega_k-\omega_l)t]\sin^2[(\omega_0+\omega_l+\omega)t/2]}{\omega_0+\omega_l+\omega}\right\rbrace.
\label{rates-modulation}
\end{multline}

Under impulsive $\pi$-flips of the qubit phase at periods $\tau$, the Floquet expansion  of $\epsilon(t)$  yields $\varepsilon_0=0$, $\varepsilon_{\pm}1=\pm2/\pi$,
$\varepsilon_{k>1}= \frac{2}{i(2k)\pi}$, and $\omega_k=(2k+1) \pi/\tau$. These are respectively the $k$th-harmonic amplitude and frequency of the periodic $\pi$-flip modulations.

Taking into account the additive contributions of cold and hot bats to $G_T(\omega)$, i.e. $G_T(\omega)=G^C_T(\omega)+G^H_T(\omega)$ and assuming the conditions of  Eq.(5) of the main text, we obtain the time-averaged partial decay rates as
\begin{eqnarray}
\nonumber
\overline{R}_e^C&=&\frac{8}{\pi}G_0^C(\omega_0- \Delta)[n^C(\omega_0- \Delta)+1],\\
\nonumber
\overline{R}_g^C&=&\frac{8}{\pi}G_0^C(\omega_0- \Delta)n^C(\omega_0- \Delta),\\
\nonumber
\overline{R}_e^H&=&\frac{8}{\pi}G_0^H(\omega_0+ \Delta)[n^H(\omega_0+ \Delta)+1],\\
\overline{R}_g^H&=&\frac{8}{\pi}G_0^H(\omega_0+ \Delta)n^H(\omega_0+ \Delta).
\label{decay-rates-limit-modulation}
\end{eqnarray}

where $\Delta=2\pi/\tau$. Under these assumptions, we can find a {\it periodic quasi-steady} state with
 constant polarization $\overline{S}_S$ 
%and obtain Eq.~\eqref{steady-state-polarization-modulation}
, namely, with the use of Eq.~\eqref{decay-rates-limit-modulation}
\begin{eqnarray}
&\overline{S}_S=- \notag \\
&\frac{G_0^C(\omega_0-\Delta)+G_0^H(\omega_0+\Delta\omega)}{2\lbrace G_0^C(\omega_0-\Delta)[2n^C(\omega_0-\Delta)+1]+G_0^H(\omega_0+\Delta)[2n^H(\omega_0+\Delta)+1]\rbrace }.
\label{eq:}
\end{eqnarray}
Using this result  and separating the $C$ and $H$ contributions,  we obtain the heat flows in the steady state 
\begin{eqnarray}
&\overline{\dot{S}^C_S}\approx
 -(\overline{R}_g^C+\overline{R}_e^C)\overline{S}_S+\frac{\overline{R}_g^C-\overline{R}_e^C}{2}= \notag \\
\nonumber
&\frac{G_0^C(\omega_0-\Delta) G_0^H(\omega_0+\Delta)[n_T^C(\omega_0-\Delta)-n_T^H(\omega_0+\Delta)]}{G_0^C(\omega_0-\Delta)[2n_T^C(\omega_0-\Delta)+1]+G_0^H(\omega_0+\Delta)[2n_T^H(\omega_0+\Delta)+1]}\\
&=-\overline{\dot{S}^H_S}.
\end{eqnarray}
For arbitrarily  wide $C$ and $H$ coupling spectra, the heat flow reads
\begin{eqnarray}
\nonumber
&&\overline{\dot{S}^{C(m)}_S}=\\
&&\frac{G_0^C(\omega_0-\Delta) G_0^H(\omega_0+\Delta)[n^C(\omega_0-\Delta)-n_T^H(\omega_0+\Delta)]}{K} \nonumber\\
\nonumber
&+&\frac{G_0^C(\omega_0+\Delta) G_0^H(\omega_0-\Delta)[n^C(\omega_0+\Delta)-n_T^H(\omega_0-\Delta)]}{K}\\
\nonumber
&+&\frac{G_0^C(\omega_0+\Delta) G_0^H(\omega_0+\Delta)[n^C(\omega_0+\Delta)-n_T^H(\omega_0+\Delta)]}{K}\\
\nonumber
&+&\frac{G_0^C(\omega_0-\Delta) G_0^H(\omega_0-\Delta)[n^C(\omega_0-\Delta)-n_T^H(\omega_0-\Delta)]}{K}
\\
&=&-\overline{\dot{S}^H_S},
\label{steady-heat-flow-markov}
\end{eqnarray}
where 
\begin{eqnarray}
\nonumber
K&=&G_0^C(\omega_0-\Delta) [2n_T^C(\omega_0-\Delta)+1] \nonumber\\
&+&G_0^C(\omega_0+\Delta) [2n_T^C(\omega_0+\Delta)+1] \nonumber\\
\nonumber
&+&G_0^H(\omega_0-\Delta) [2n_T^H(\omega_0-\Delta)+1]\\
&+&G_0^H(\omega_0+\Delta) [2n_T^H(\omega_0+\Delta)+1].
\label{K}
\end{eqnarray}
In the case of flat and wide (Markovian-like) spectra, all four terms are of comparable magnitude, hence $\overline{\dot{S}^C_S}$ is negative, i.e. there is no QR action. By contrast, under the  assumption of Eq.(5) of the main test, the last three terms are negligible, yielding Eq.(7) and possible change of sign of the heat flow for QR.

\subsection*{Steady states and currents by Floquet expansion of the Lindblad operator}

An alternative method for performing the Floquet expansion of the steady-state of a periodiaclly-flipped qubit coupled to two baths is based on the Lindblad operator, as expounded in the tutorial in Ref. \cite{tutorial} 

For the qubit density  $\tilde{\rho}$ to be a steady state of the Lindblad operator we expand $\mathscr{\mathcal{L=\sum L}}_{q}^{j}$ where  $j=H,C$, and q is the Floquet harmonic. The expansion yields (see notation in the main text)

\begin{gather}
\mathcal{L}_{q}^{j}\rho=\frac{P(q)}{2}\Bigl(G_{j}(\omega_{0}+q\Delta)\bigl([\sigma^{-}\rho,\sigma^{+}]+[\sigma^{-},\rho\sigma^{+}]\bigr)+\notag \\
G_{j}(-\omega_{0}-q\Delta)\bigl([\sigma^{+}\rho,\sigma^{-}]+[\sigma^{+},\rho\sigma^{-}]\bigr)\Bigr)\label{gen_qubit2}
\end{gather}

The qubit steady-state then has the form 

 $\tilde{\rho}=\left(\begin{array}{cc}
A & 0\\
0 & B
\end{array}\right)$ 

\begin{equation}
\frac{A}{B}=\frac{\sum_{q,j}P(q)G^{j}(\omega_{0}+q\Delta)e^{-\frac{\omega_{0}+q\Delta}{T_{j}}}}{\sum_{q,j}P(q)G^{j}(\omega_{0}+q\Delta)}
\label{eq:ss}
\end{equation}

where $\Delta=\frac{2\pi}{\tau}$. The cold (hot) current is then given by

\begin{equation}
J_{C(H)}=\sum_{q}P(q)(\omega_{0}+q\Delta)G^{C(H)}(\omega_{0}+q\Delta)\frac{e^{-\frac{(\omega_{0}+q\Delta)}{T_{C(H)}}}-\frac{A}{B}}{\frac{A}{B}+1}
\label{eq:curr}
\end{equation}

 The magnitudes and signs of these steady-state currents are the same as for the time-averaged ME solutions detailed above.

\begin{thebibliography}{99}
\bibitem{nernst} Nachr. Kgl. Ges. Wiss. Goett. \textbf{1}, 1 (1906);
Kgl. Pr. Akad. Wiss. \textbf{52}, 933 (1906);
W. Nernst, Sitzberg. Preuss. Akad. Wiss. Phys.-Math. p. 134 (1912), \textit{The New Heat Theorem} (Dutter, New York, 1926).
\bibitem{landsberg} P.T. Landsberg, Rev. Mod. Phys. \textbf{28}, 363 (1956).
\bibitem{belgiorno} F. Belgiorno, J. Phys. A \textbf{36}, 8195 (2003).
\bibitem{gordon-jap} E. Geva and R. Kosloff, J. Chem. Phys. \textbf{96}, 3054 (1992);
T. Feldmann et al., Am. J. Phys. \textbf{64}, 4 (1996);
Y. Rezek and R. Kosloff, New J. Phys. \textbf{8}, 83 (2006);
R. Kosloff et al., J. of Appl. Phys. \textbf{87}, 11 (2000);
Y.Rezek and R. Kosloff, New J. Phys. \textbf{8}, 83 (2006);
T. Feldmann and R. Kosloff, EPL \textbf{89}, 20004 (2010).
\bibitem{segal-pre} D. Segal and A. Nitzan, Phys. Rev. E \textbf{73}, 026109 (2006);
D. Segal, Phys. Rev. Lett. \textbf{101}, 260601 (2008).
\bibitem{linden} N. Linden and S. Popescu and P. Skrzypczyk,  et.al., Phys. Rev. Lett. \textbf{105}, 130401 (2010).
\bibitem{kosloffprl12} A. Levy and R. Kosloff, Phys. Rev. Lett. \textbf{108}, 070604 (2012);
B. Cleuren et al. ibid, 120603 (2012); 
A. Mari and J. Eisert, ibid 120602(2012).
\bibitem{SchulzePRA10}
\bibinfo{author}{R. Schulze}, \bibinfo{author}{C. Genes} \&
  \bibinfo{author}{H. Ritsch,}
\newblock \emph{\bibinfo{journal}{Phys Rev. A}} \textbf{\bibinfo{volume}{81}},
  \bibinfo{pages}{063820} (\bibinfo{year}{2010});
\bibinfo{author}{X. Chen~et al.,}
\newblock \emph{\bibinfo{journal}{Phys. Rev. Lett.}}
  \textbf{\bibinfo{volume}{104}}, \bibinfo{pages}{063002}
  (\bibinfo{year}{2010}).
\bibinfo{author}{A. Ruschhaupt,} \& \bibinfo{author}{J.~G. Muga, }
\newblock \emph{\bibinfo{journal}{Phys. Rev. A}} \textbf{\bibinfo{volume}{70}},
  \bibinfo{pages}{061604(R)} (\bibinfo{year}{2004}).
\bibinfo{author}{A. Ruschhaupt,}, \bibinfo{author}{J.~G. Muga,} \&
  \bibinfo{author}{M.~G. Raizen, }
\newblock \emph{\bibinfo{journal}{J. Phys. B}}
  \textbf{\bibinfo{volume}{39}}, \bibinfo{pages}{3833} (\bibinfo{year}{2006});
  J. Birjukov et al. Eur. Phys. J. B \textbf{64}, 105 (2008).
\bibitem{alicki} R. Alicki, Jour. Phys. A \textbf{12}, L103 (1979); 
R. Alicki et al. Open Sys. \& Inf. Dyn. \textbf{11}, 205 (2004); 
J. Gemmer, M. Michel and G. Mahler, \textit{Quantum Thermodynamics}. (Springer, 2010);
\bibinfo{author}{A.~E. Allahverdyan, } \& \bibinfo{author}{T.~M. Nieuwenhuizen, }
\newblock \emph{\bibinfo{journal}{Phys. Rev. E}} \textbf{\bibinfo{volume}{71}},
  \bibinfo{pages}{046107} (\bibinfo{year}{2005}).
\bibitem{pop} E. Pop, S. Sinha, and K.E. Goodson, Proc. IEEE, \textbf{94(8)} 1587 (2006)
\bibitem{kittel} C. Kittel, \textit{Quantum Theory of Solids}. (Wiley, 1987).
\bibitem{Alexander}S. Alexander, O. Entin-Wohlman and R. Orbach, Phys. Rev. B \textbf{32}, 6447 (1985); M.P. Fontana et al., Phil. Mag. B \textbf{65}, 143 (1992)
\bibitem{R1}
P. Pringsheim, Z. Phys. \textbf{57}, 739 (1929);
 L. Landau, J. Phys. (Moscow) \textbf{10}, 503 (1946);
 G.C. Dousmanis, C.W. Mueller, H. Nelson, and K.G. Petzinger; Phys. Rev. \textbf{133}, A316 (1964);
N. Djeu and W.T. Whitney,  Phys. Rev. Lett. \textbf{46}, 236 (1981);
S.A. Egorov and J.L. Skinner, J. Chem. Phys. \textbf{103}, 1533 (1995);
S. Lloyd,  Phys. Rev. A \textbf{56}, 3374 (1997);
C.E. Mungan and T.R. Gosnell,  Advances in Atomic, Molecular, and Optical Physics, edited by B. Bederson and H. Walther (Academic Press, San Diego, 1999) Vol. 40, p. 161;
R.I. Epstein, M.I. Buchwald, B.C. Edwards, T.R. Gosnell, and C.E. Mungan, Nature \textbf{377}, 500 (1995);
 C.E. Mungan, M.I. Buchwald, B.C. Edwards, R.I. Epstein, and T.R. Gosnell,  Phys. Rev. Lett. \textbf{78}, 1030 (1997);
C.E. Mungan,  J. Opt. Soc. Am. B \textbf{20}, 1075 (2003).
\bibitem{kurizki-prl}  A.G. Kofman and G. Kurizki, Phys. Rev. Lett. \textbf{93}, 13 (2004);
G. Gordon, N. Erez and G. Kurizki J. Phys. B \textbf{10}, S75 (2007);
G. Gordon, G. Kurizki and D.A. Lidar, Phys. Rev. Lett. \textbf{101}, 010403 (2008);
\bibitem {Almog} I. Almog et al., J. Phys B \textbf{44}, 154006 (2011).
\bibitem{alvarez} G. Alvarez et al., Phys. Rev. Lett. \textbf{105}, 160401 (2010):
\bibitem{noam-nature} N. Erez et al., Nature \textbf{452}, 724 (2008); G. Gordon et al., NJP \textbf{11}, 123025 (2009).
NJP \textbf{12}, 053033 (2010);
T. Jahnke and G. Mahler, Europhys. Lett. \textbf{90}, 50008 (2010).
D. D. B. Rao and G. Kurizki, Phys. Rev. A \textbf{83} 032105 (2011).
\bibitem{17} S.T. Smith and R. Onofrio, Eur Phys J. \textbf{B61}, 271 (2008);
H. Hasegawa, Phys Rev. \textbf{E83}, 021104 (2011);
A. Carcaterra and A. Akay, Phys Rev. \textbf{E84}, 011121 (2011).
\bibitem{tutorial}R. Alicki  et al. 		arXiv:1205.4552v1 [quant-ph] (2012).
\bibitem{alilid} R. Alicki et al. Phys. Rev. A \textbf{73}, 052311 (2006).
\bibitem{Kurizki:2001} A.G. Kofman and G. Kurizki, Phys. Rev. Lett. {\bf 87}, 270405, (2001);
Nature \textbf{405}, 546 (2000).
\bibitem{lindblad} G. Lindblad, \textit{Non-Equilibrium Entropy and Irreversibility}. (D. Riedel, Holland 1983).
\bibitem{reichl} L.E. Reichl. \textit{A Modern Course in Statistical Physics.} (Wiley-Interscience, USA, 1998).
\bibitem{jens} J. Clausen and G. Bensky and G. Kurizki et.al., Phys. Rev. Lett. \textbf{104}, 040401 (2010);
Phys Rev A \textbf{85}, 052105(2012).
\bibitem{arxivekos} A. Levy and R. Alicki and R. Kosloff 	Phys. Rev. E \textbf{85}, 061126 (2012). 
\bibitem{NeumannSCI08}
\bibinfo{author}{P. Neumann~et al.,}
\newblock \emph{\bibinfo{journal}{Science}} \textbf{\bibinfo{volume}{320}},
  \bibinfo{pages}{1326} (\bibinfo{year}{2008}).
\bibitem{diehl} S. Diehl et al. Nature Physics, \textbf{4(11)} 878 (2008).
\end{thebibliography}
\end{document}